# Hexapole-Oriented Asymmetric-Top Molecules and Their Stereodirectional Photodissociation Dynamics

Authors: Masaaki Nakamura [a], Shiun-Jr. Yang [a], Po-Yu Tsai [a, b], Toshio Kasai [a], King-Chuen Lin [a, c, *], Dock-Chil Che [d], Andrea Lombardi [e], Federico Palazzetti [e, f, †], Vincenzo Aquilanti [e, g, h]

[a] National Taiwan University – Department of Chemistry – Taipei, Taiwan.

[b] National Chung-Hsing University – Department of Chemistry – Taichung, Taiwan.

[c] Institute of Atomic and Molecular Sciences – Taipei, Taiwan.

[d] Osaka University – Graduate School of Science, Department of Chemistry – Osaka, Japan.

[e] Università di Perugia – Dipartimento di Chimica, Biologia e Biotecnologie – Perugia, Italy

[f] Scuola Normale Superiore – Pisa, Italy.

[g] Consiglio Nazionale delle Ricerche - Istituto di Struttura della Materia, 00016 Roma, Italy

[h] Universidade Federal da Bahia – Instituto de Fìsica – Salvador, Brazil.

[*] Corresponding author: kclin@ntu.edu.tw; +886 2 33661162

[†] Corresponding author: federico.palazzetti@unipg.it; +39 075 5855521

# Abstract.

Molecular orientation is a fundamental requisite in the study of stereodirected dynamics of collisional and photoinitiated processes. In this last decade, variable hexapolar electric filters have been developed and employed for the rotational-state selection and the alignment of molecules of increasing complexity, for which main difficulties are their mass, their low symmetry and the very dense rotational manifold. In this work, for the first time, a complex molecule such as 2-bromobutane, an asymmetric-top containing a heavy atom (the bromine), has been successfully oriented by a weak homogeneous field placed downstream the hexapolar filter. Efficiency of the orientation has been characterized experimentally, by combining time-of-flight measurements and a slice-ion-imaging detection technique. The application is described to the photodissociation dynamics of the oriented 2-bromobutane, which was carried out at a laser wavelength of 234 nm, corresponding to the breaking of the C – Br bond. The Br photofragment is produced in both the ground Br ($^2P_{3/2}$) and the excited Br ($^2P_{1/2}$) electronic states and both channels are studied by the slice imaging technique, revealing new features in the velocity and angular distributions with respect to previous investigations on non-oriented molecules.

# 1. Introduction.

The importance of the steric effect in elementary chemical processes has been motivating the development of techniques that permit the achievement of control of molecular spatial, translational and internal degrees of freedom. Control of molecular velocity and directionality, through the improvement of techniques such as time-of-flight mass-spectrometry and mechanical velocity selectors, was the first important achievement, allowing one to study the effects deriving from single-collision phenomena (see for example [1]). Alignment and orientation, which consist in non-statistical distributions of the rotational angular momentum with respect a quantization axis, permit the study of spatial aspects of molecular dynamics. The alignment [2,3] indicates the polarization of the direction of the angular momentum vector, while the orientation [4] concerns both direction and sense. Their control permits to obtain information on geometrical features of molecules and spatial aspects of dynamics [5,6], otherwise concealed in free rotating molecules; separation of enantiomeric effects in photodissociation [7,8] and in collisional processes [9] are among the most intriguing features expected to be associated to the molecular orientation.

Angular distribution of photofragments can be predicted by vector correlation of intrinsic molecular properties (*e. g.* permanent dipole moment and transition dipole moment) and external vectors defined in the laboratory-frame (*e. g.* laser polarizations and orientation vector) [10-13]. Hexapolar electric fields are used to align and select rotational states of molecules with a permanent dipole moment. This is the technique of choice in this paper. Hexapoles were initially employed for the rotational state-selection of linear [14,15] and symmetric-top molecules [16-18], and more recently on asymmetric-tops [19,20] and in the selection of conformers of the asymmetric-top 2-butanol [21] (for studies on dimers see [22,23], for other applications of the hexapolar filters see for example [24]). Orientation is achieved by applying homogenous electric fields to hexapole-aligned molecules [25], as implemented here.

Other techniques to obtain molecular orientation make use of strong homogenous electric fields (brute-force techniques) [26-28], laser absorption techniques [29-30] and non-resonant intense laser techniques [31-32]. Natural alignment in supersonic molecular beams, by effect of the collisions of seeding gases, was experimentally proven for linear [33] and disk-shaped molecules [34] and exploited for molecular-beam scattering studies of the anisotropy of intermolecular forces (see also References [35-36]).

In the present work, the asymmetric-top (chiral) molecule 2-bromobutane [37] is oriented through a 70 cm-length hexapole. 2-bromobutane is a molecule with a highly dense manifold of rotational states; it presents three rotamers, T, G+ and G- (Figure 1) whose Ray's parameter [19], an estimate of the symmetry degree of the molecule (it varies from -1 to +1, with the limits denoting the prolate- and oblate-tops respectively), is far from the prolate-top limit: -0.91 for T, -0.41 for G+ and -0.76 for G-.

Time-of-flight measurements combined with the slice-ion-imaging technique permit the detection of the molecular orientation: the orienting electric field axis is parallel to time-of-flight axis, thus the photofragments have a different time-of-flight arrival time depending on the forward or backward displacement of the bromine atom. In the case of 2-bromobutane, it was observed that the bromine atom is oriented toward the negative pole of the orienting field. The ion-imaging technique can also serve to verify the orientation by means of a strong up-down asymmetry, not manifested in the absence of molecular orientation (see ion images in the photodissociation of the oriented "nearly linear-top" molecule OCS by Rakitzis et al. [11]). The hexapole turns out to be an efficient technique, even for molecules far from the symmetric-top limit and with a populated and very dense rotational manifold. Convenient is also the finding that the molecular orientation is achieved by using weak (100~200 V/cm) orienting fields.

Photodissociation of the oriented 2-bromobutane is studied by a 234 nm laser, in order to detect the Br photofragment, by slice-ion-imaging technique. The excitation at 234nm, namely the A band, results in correspondence of the C-Br bond-breaking, emitting a Br atom photofragment, for which there are two accessible fine-structure states: the ground state Br ($^2P_{3/2}$) indicated by Br and the excited state Br ($^2P_{1/2}$) indicated by Br$^*$. Both photofragments can be observed separately by resonance-enhanced multiphoton ionization (REMPI). Recent studies with non-oriented molecules have reported [38, 39] that these two states show independent photofragment distributions, and report two velocity components for each state. Regarding the faster components, both velocity and angular distributions are characterized by our results, while the slower components were not observed as such in our experiment, but slow to arise to the overlapped signals in the time domain, originated by the two isotopes of bromine atom, $^{79}$Br and $^{81}$Br, occurring in nature with approximately equal abundance; in the present experiment the two signals are almost isolated on sliced images by using the weak ion extraction field.

The article is structured as follows: in Section 2, we describe the experimental apparatus for the slice-ion-imaging technique and the time-of-flight measurements; in Section 3, we discuss the results related to the hexapole orientation and the photodissociation of 2-bromobutane; in Section 4, finally, in Section 4, we finish with the conclusions.

## 2. The experimental methods.

### 2.1 Slice ion-imaging technique.

The experimental setup for the slice ion-imaging is illustrated in Fig.2. The vacuum system is composed of three chambers, separated by differential pumping: they accommodate the molecular-beam source, the hexapole selector and the imaging detector system. The hexapole and the detector chambers are kept under high-vacuum condition at a working pressure of $5.0 \times 10^{-7}$ torr. The 2-bromobutane is used as purchased; its vapor pressure is 65 torr at room temperature. It is ejected through a pulsed valve (General Valve Co.) with a 0.6 mm-diameter orifice and expanded into the source chamber. After passing through a 2 mm-diameter skimmer located at 50 mm distance in the downstream, the skimmed beam is collimated with a 5 mm-diameter collimator, located in the hexapole chamber, placed 60 mm away from the skimmer. The molecular beam passes through the hexapole chamber and reaches the ion lens, where it is intersected perpendicularly by a linearly polarized laser.

The dye laser beam (Lambda Physik PD 3000) is pumped by a 308 nm XeCl excimer laser (Lambda Physik LPX 200) and operated synchronously with the pulsed valve at 10 Hz. Its output is frequency doubled to emit at *ca.*234 nm with 400 μJ per pulse. Its polarization is fixed perpendicularly to the time-of-flight axis and parallel to the imaging surface. A one-color laser experiment is conducted to photolyze the 2-bromobutane molecule, followed by probe either of the dissociated fine-structure ground $Br(^2P_{3/2})$ state at 233.7 nm and of the excited $Br(^2P_{1/2})$ state at 234.0 nm, using a (2+1) resonance-enhanced multiphoton ionization (REMPI) technique. The generated ion cloud is accelerated toward the micro-channel plate (MCP) detector (Galileo FM3040) by an extraction field and is projected onto the phosphor screen. The ion lens comprises four hollow metal disks according to the "three-electrode" design [40]. Such a design can stretch the photofragment ion cloud along the time-of-flight axis, while keeping velocity mapping condition fulfilled on the detector screen. For the heavy bromine ion, the head-to-tail arrival time of the ion cloud is spread to 250 ns in the relatively short flight distance of about 30 cm. The MCP detector is gated with a high-voltage pulse (Kentech Instruments Ltd.) to take an image only of the center slice of the ion cloud. The sliced images are centered on the isotope $^{79}$Br (the other stable isotope is $^{81}$Br). The gate pulse duration is controlled at 25 ns, thus clipping 10 % of the whole ion cloud sphere. The 2D ion images are taken by a charge-coupled device camera (Pixefly 200XL4078). Meanwhile, the signal intensity is monitored with a photomultiplier tube (PMT) simultaneously. The images are accumulated for 140,000 ~ 150,000 laser shots to obtain a better signal-to-noise ratio. The same numbers of images collected at off resonance are subtracted to eliminate the unwanted contributions from background gases and clusters.

**2.2. The hexapole and the time-of-flight measurements.**

The efficiency of the molecular orientation is monitored by time-of-flight measurements of the photofragments. In case of non-oriented molecules, the population of fast and slow peaks is equally distributed. In contrast, for oriented molecules, the two peaks appear at different intensities because of the induced biased populations. As illustrated in Fig. 2, the hexapole state selector, installed in the hexapole chamber, is used to select specific rotational states of a molecular beam by means of Stark effect interaction. It is composed of six stainless steel rods each of 4 mm-diameter and 70 cm-length. The minimum distance between the hexapole axis and the surface of the rods, the hexapole radius, is 4 mm. The hexapole exit is at 2cm from the entrance of ion lens. This small distance permits the adiabatic passage of the hexapolar state-selected molecules to the orienting field, *i. e.* without losing alignment. In the current apparatus, we have not included the beam stop, which is normally used to block the direct beam, in order to obtain a non-zero beam intensity in absence of hexapole electric field. As mentioned in the introduction, the hexapole operates a rotational state-selection resulting in a non-statistical distribution of the molecular populations, that are eventually focused onto the ion lens The extraction field, set at *ca.* 200 V/cm (in the ion lens) plays two roles: (i) it orients the molecules and (ii) it accelerates the photofragment to the detector. 2-Bromobutane is photodissociated at a laser beam polarized along the time-of-flight axis; the forward and backward Br atom fragments are detected using (2+1) REMPI technique. The time-of-flight signal is acquired from the micro channel plate, MCP, directly, instead of using a photo multiplier tube, PMT, to recall the phosphor screen output. The MCP is kept opened without any gating in the measurement, in order to observe a whole ion cloud.

## 3. Results and discussions.

### 3.1. Hexapolar state-selection and molecular orientation.

In the hexapole (see Section 2.2) an inhomogeneous electric field is generated that has zero intensity along the hexapole axis and increases by moving toward the rods. (For theoretical background of the hexapolar technique see [20, 41, 42]). The effect of the electric field is to focus the molecular beam along its axis, aligning and selecting the molecules according to their rotational states. The intensity of the electric field $E$ depends on the distance $r$ from the axis of the hexapole and is given by

$$E = 3V_0 \frac{r^2}{R^3}, \quad (1)$$

where $R$ is the hexapole radius (see Section 2.2) and $V_0$ is the voltage applied to the rods of the hexapole.

The "focusing curve" is a measure of the variation of the beam intensity as a function of the hexapole voltage. Its shape mainly depends on the components of the permanent dipole moment [12] along the principal axis of inertia. Figure 3 shows the focusing curve of the pure beam of 2-bromobutane: the beam intensity slightly increases from the initial value, set at zero at 0 kV up to *ca.* 1.5 kV, then the enhancement of the signal is much stronger and continuously increases until the maximum applied voltage of 5 kV. The focusing effect of the hexapole increases the beam intensity up to a factor two. Since no beam stop is included (see Section 2.2), the non-zero intensity has been subtracted (see Supporting Information for a detail of the rotational selected-states). The velocity distribution of the supersonic molecular beam is given by the following equation:

$$F_v(v) = v^3 exp(-\frac{(v-v_M)^2}{\alpha_M^2}) \quad (2)$$

under our typical operating conditions, we found $v_M$ = 450 m/s (the peak velocity) and $\alpha_M$ = 50 m/s (the full-width half-maximum). Simulation of the molecular trajectories based on the present experimental apparatus was performed to reproduce the experimental focusing curve and calculate the rotational state-selection. The simulated trajectories are assumed starting from the nozzle, according to a Gaussian distribution. The maximum value of the total angular momentum $J$ is 22, as established according to a convergence criterion. The rotational temperature inferred from the comparison between simulated and experimental focusing curves is 30 K. Because of the operating conditions, a pure beam with a low stagnation pressure (see Section 2.1), we assume a vibrational temperature of 300 K. The three stable conformers of 2-bromobutane, T, G+ and G- (relevant physical properties of the conformers, calculated at HF/6-311+G (d, p) level of theory, are reported in Table 1, see also Reference [37]) have been considered in the simulations and the total focusing curve has been built by summing the focusing curve of each conformer, according to the relative populations calculated by the Boltzmann distribution function at the assumed vibrational temperature [21]. Differently from the case of linear-tops, symmetric-tops and light nearly symmetric-top molecules, the very dense manifold of the rotational levels does not permit to select a single rotational state selection. However, the hexapole state-selection generates a non-statistical distribution of the rotational states, which can be determined by a best-fit comparison between experimental and simulated focusing curves. Complementarily to the Ray's parameter (see Table 1), the deformation indices [43, 44] $\xi_+$ and $\xi_-$ (Table 2) are a measure of the molecular shape

$$\xi_+ = \frac{\xi_2^2 - \xi_3^2}{\rho^2}$$
$$\xi_- = \frac{\xi_2^2 - \xi_1^2}{\rho^2} \qquad (3)$$

and are expressed by the kinematic invariants $\xi_1$, $\xi_2$, $\xi_3$, and the hyperradius $\rho$,

$$\xi_1^2 = \frac{I_C - I_A + I_B}{2M_N}$$
$$\xi_2^2 = \frac{I_A - I_B + I_C}{2M_N}$$
$$\xi_3^2 = \frac{I_A + I_B - I_C}{2M_N} \qquad (4)$$
$$\rho^2 = \xi_1^2 + \xi_2^2 + \xi_3^2,$$

where $I_A$, $I_B$ and $I_C$ are the components of the moments of inertia, with $I_A \leq I_B \leq I_C$ and $M_N$ is the mass of the molecule (for 2-bromobutane, it is 137.02 a. m. u.). The deformation index $\xi_+$ is a non-negative number and is 0 for prolate-top molecules, its value is 0.15, 0.31 and 0.17 for T, G+ and G- conformers, respectively. The $\xi_-$ deformation index is instead a non-positive number and is 0 for oblate-tops, its value is in this case -0.63, -0.28 and -0.43 for the three conformers.

Molecular orientation (See Section 2.2 and Figure 2) is analyzed by time-of-flight measurements, according to the following considerations: the bromine photofragment is recoiled forward, along the drift tube [40], and reaches the detector directly; the backward-recoiled fragment flies toward the repeller plate [40] prior to being accelerated back to the detector by the extraction field, and arrives at the detector later with respect to the forward-recoiled photofragment. As discussed in the previous section, randomly oriented molecules do not manifest such a different intensity, because the forward and backward-recoiled fragments have the same population.

Fig. 4(a) shows the time-of-flight of Br* measured by REMPI technique at 234.0 nm, with the laser polarization parallel to the time-of-flight axis. As discussed later in Section 3.2, Br* has a large anisotropy parameter ($\beta = 1.85$), in this case the photofragments are distributed along the laser polarization axis (parallel to the time-of-flight axis) and the recoiled photofragments yield two peaks along the time-of-flight axis. The earliest peak corresponds to $^{79}$Br, the latest one corresponds to $^{81}$Br, while the central peak is caused by the overlap between the two ion cloud spheres of the $^{79}$Br and $^{81}$Br isotopes. When the hexapole voltage is not applied, the molecules are not rotationally selected and their orientation is random.

As shown in Fig 4(a), this behavior is represented by a black curve, where the intensities of the earliest and latest peaks are very similar. By increasing the voltage to 4 kV (red line in the figure), the earliest peaks are largely enhanced, whereas the latest one does not change significantly. Figures 4 (b1) and (b2) show the simulation of the experimental data reported in Fig. 4(a), when the hexapole voltage is off (Fig. 4 (b1)) and on (Fig. 4 (b2)); the deconvolution of the central peak is also reported. As shown in Fig 4 (b1), $^{79}$Br (red dashed line) and $^{81}$Br (blue dotted line) give two peaks with similar intensities when the hexapole voltage is off, and the central peak is the sum of the two peaks (green solid line).

Figure 4 (b2) shows the deconvolution of the signal measured when the hexapole voltage is applied. In this case, the deconvolution requires consideration of the orientation probability distribution function, OPDF, as a function of $\theta$, the angle between the permanent dipole moment vector and the orientation axis that can be expressed as a linear combination of Legendre polynomials $P_n$:

$$\text{OPDF}(\cos\theta) = \sum_{k=0}^{2J} c_k P_k(\cos\theta). \qquad (5)$$

Up to three terms have been considered, obtaining the coefficients $c_0 = 0.5$, $c_1 = 0.35$ and $c_2 = 0.1$. In spite of the simplification, with satisfactory fit of the experimental results.

The ion lens setup described in Section 2.1, produces a weak static electric field, *ca.* 200 V/cm, that acts as ion extractor and permits the molecular orientation, as shown in Fig. 4. With the current setup, it is not possible to change intensity and direction of the orienting D. C. (Direct Current) field. To overcome this drawback, a pulsed voltage is added in the extraction stage to switch between "orientation mode" and "extraction mode". Since ion extraction starts after the intersection of the molecular beam and the photolysis/ionization laser, it is possible to orient molecules and to extract photofragment ions separately on the time axis [12].

As shown in Fig.5(a), the backward peaks of both isotopes are enhanced by the pulsed-voltage addition (red line), with respect to the forward peak enhancement by the fixed orientation field (black line). Note that the signal intensity oscillations were caused by the laser power fluctuation. The pulsed field is applied at the first extraction stage in which the molecular beam and the photolysis laser were intersected to flip over the direction of the orientation of 2-bromobutane, according to the forward-backward displacement of the Br atom along the time-of-flight axis.

The time scheme of the first extraction stage is described in Fig. 5(b). The applied pulse rises when the molecular beam reaches the repeller plate and lasts for about 28 μs until right before the intersection with the photolysis laser. In such a way, the pulsed field induces a molecular backward orientation, we call this stage "orientation mode". By removing the pulsed voltage, the electric field works as an extractor: we call this stage "extraction mode". The response time of the pulsed voltage is very short, only 30 ns, avoiding alteration of the peak appearance along the time-of-flight axis. It means that the pulsed voltage does not interfere with the ion extraction. To flip over the orientation, a pulsed field of *ca.* 200 V/cm and opposite polarity with respect to the extraction field is applied. In the current setup, the strength of the pulse can be increased up to 1.3 kV/cm for the backward orientation (same as the red line in Fig. 5(a)) and 2kV/cm for forward orientation (same as the black line in Fig. 5(a)). By decreasing the distance between the repeller plate (R) and the first extractor lens plate (L1) the orienting field intensity can be further increased. This is important for future applications; generally, a stronger orientation field is required to orient asymmetric top molecules, in contrast with the weak extraction field, required for sliced imaging. Use of the pulsed field option represents an optimal solution.

**3.2. Photodissociation dynamics of 2-bromobutane investigated by slice-ion-imaging.**

The hexapole-oriented 2-bromobutane has been photodissociated at a laser wavelength of *ca.* 234 nm and the resultant Br and Br* have been ionized by (2+1) REMPI at 233.7 and 234.0 nm, respectively. Their sliced ion images, shown in Fig. 6, reveal information on both velocities and angular distributions. The photolysis laser is polarized perpendicular to the drift tube and parallel to the detector plane, which corresponds to the vertical direction on the images. The analysis of image results shows a maximum velocity of 965 and 920 m/s for Br and Br*, equivalent to a center-of-mass total translational energy disposal of 87.8 and 79.8 kJ/mol, respectively, while the corresponding anisotropy parameter $\beta$ is 1.49±0.03 for Br and 1.85±0.07 for Br*.

Our results agree substantially with those obtained at the same wavelength by Zhou *et al.*[38]. It is however of fundamental importance to note that in Zhou *et al.* experiment, the ion imaging yields an inner and an outer ring for both Br and Br* atoms, corresponding to two velocity components. They interpreted the faster component as a result from the excited-state dissociation, while the slower one was ascribed to the dissociation on the ground state surface via internal conversion. While the outer rings (the faster component) turn out to be consistent with our results, we do not observe the inner rings (the slow component) for both the ground or the excited states, except for a small region close to the center: it appears that the slow component stems from the partial overlap between two stable isotopes, $^{79}$Br and $^{81}$Br, whose natural abundance is of 50.7 and 49.3 %, respectively. Under the current experimental conditions, these two isotopes cannot be completely separated within the time-of-fight arrival time; in fact, although the slicing gate pulse is positioned to detect the $^{79}$Br isotope, traces of $^{81}$Br are still observed. While taking a series of time slices of both isotopes by shifting the pulsed gate timing, it turned out that the later half-sphere of the $^{79}$Br ion cloud overlaps with the earlier half-sphere of $^{81}$Br, thereby confirming that the inner region is due to the presence of $^{81}$Br. Br has a slightly more intense "slow region" than Br*, which is ascribed to the smaller anisotropy parameter carried by Br, resulting in a larger overlap of the equatorial regions of the two ion cloud spheres. In addition, a smaller size of the Br* ion sphere yields a smaller overlap.

Figure 7a and 7b show the experimental and simulated velocity mapping imaging outcome of the Br* photofragments dissociated from the oriented 2-bromobutane. Differently from the above-mentioned sliced image, a longer gate width of about 400 ns has been used to cover the whole ion cloud spheres; thus, both $^{79}$Br and $^{81}$Br are fully superimposed on the image. The laser polarization is rotated 45° toward the MCP detector surface [10, 11]. The asymmetric image and subsequent angular distribution are then obtained (Figure 7c). To fit the angular distribution of the outer ring, we used the following equation:

$$I(\theta) = 1 + \beta_1 P_1(\cos\theta) + \beta_2 P_2(\cos\theta), \quad (6)$$

where $P_1$ and $P_2$ are the first and the second terms of Legendre polynomials $P_n$. while $\beta_1$ and $\beta_2$ are the expansion coefficient for $P_1$ and $P_2$, respectively. The best-fit to the experimental results yields $\beta_1 = 0.53$ and $\beta_2 = 1.56$ for the Br* fragment. $\beta_1$ represents the up-down asymmetry of the outer ring, assuming value different from zero only for oriented molecules; it is associated with the angles $\alpha$ and $\chi$, where $\alpha$ is the angle between the permanent dipole moment **d** and the fragment recoil velocity **v** and $\chi$ is the angle between the transition dipole moment **μ** and **v**. $\beta_1$ vanishes when $\alpha$ is 0° or 180° or when $\chi$ is 0° or 90° or 180°. The angles $\alpha$ and $\chi$ are fundamental for the accurate characterization of the fragment spatial behavior in the photodissociation dynamics of polyatomic molecules. They have been obtained by simulation in order to compute the 2D projection of the 3D photofragment angular distribution of oriented molecules (Fig. 7b). We consider only a single electronic state, being weak in the diabatic coupling with higher ones. Rakitzis et al. [10] give the photofragment angular distribution in a molecular frame as

$$I(\gamma, \delta, \varphi_{O\epsilon}) = [1 + 2P_2(\cos\chi)P_2(\cos\gamma)][1 + 2c_1 \cos\alpha \cos\delta + 2c_2 P_2(\cos\alpha)P_2(\cos\delta)] +$$
$$6\sin\gamma\cos\gamma\sin\chi\cos\chi\sin\alpha\sin\delta\cos(\varphi_{O\varepsilon} + \varphi_{\mu d})[c_1 + 3c_2\cos\alpha\cos\delta] +$$
$$\frac{9}{8}c_2\sin^2\gamma\sin^2\chi\sin^2\alpha\sin^2\delta\cos 2(\varphi_{O\varepsilon} + \varphi_{\mu d}) \quad (7)$$

where the angles $\gamma$, $\delta$, $\varphi_{O\varepsilon}$, $\alpha$, $\chi$ and $\varphi_{\mu d}$ are defined in the molecular frame $xyz$ (Figure 7d). The $z$-axis corresponds to the fragment recoil velocity **v** and the $xz$-plane corresponds to the surface formed by **v** and the laser polarization **ε** (Figure 7d); $\gamma$ is the angle between **v** and **ε**, $\delta$ is included between **v** and the direction of the orienting field **O**, $\varphi_{O\epsilon}$ is defined between **O** and **ε**, finally $\varphi_{\mu d}$ is the angle included between **μ** and **d**. These angles can be expressed in terms of laboratory frame angles $XYZ$ by the following relations:

$$\cos\gamma = \cos\Omega\cos\Gamma + \sin\Omega\sin\Gamma\cos\Theta, \quad (8a)$$

$$\cos\delta = \cos\Omega\cos\Delta + \sin\Omega\sin\Delta\cos(\Theta - \Phi), \quad (8b)$$

$$\cos\varphi_{O\varepsilon} = \{\sin^2\Omega \cos\Gamma \cos\Delta + \sin\Gamma \sin\Delta \cos\Phi -$$
$$\sin\Omega \cos\Omega [\sin\Delta \cos\Gamma \cos(\Phi - \Theta) + \sin\Gamma \cos\Delta \cos\Theta] -$$
$$\sin^2\Omega \sin\Gamma \sin\Delta \cos(\Phi - \Theta) \cos\Theta\}/(\sin\gamma \sin\delta), \quad (8c)$$

$$\sin\varphi_{O\varepsilon} = \{\cos\Omega \sin\Gamma \sin\Delta \sin\Phi - \sin\Omega [\sin\Delta \cos\Gamma \sin(\Phi - \Theta) +$$
$$\sin\Gamma \cos\Delta \sin\Theta]\}/(\sin\gamma \sin\delta), \quad (8d)$$

where the Z-axis of the laboratory frame corresponds to the direction of the time-of-flight axis. The photolysis laser light propagates along the Y-axis, and its laser polarization ε rotates in XZ-plane with a polar angle $\Gamma$ about the Z-axis. The angles $\Omega$ and $\Theta$ are, respectively, the elevation and azimuth that define **v** with respect to the Z-axis; while the angles $\Delta$ and $\Phi$ are the elevation and azimuth that define **O** (Figure 7e). Substituting Equations (8a) - (8d) into Equation (7) yields a photofragment angular distribution in the laboratory frame, where the experimental conditions are fixed at $\Gamma = 45°$, $\Delta = 180°$ and $\Phi = 0°$. The resulting large anisotropy parameter ($\beta = 1.85\pm0.07$) implies that the photofragment Br* recoil velocity is almost parallel with the transition dipole moment. Therefore, the contribution of $\varphi_{\mu d}$ is negligible and can be set at 0°. Given $c_1 = 0.35$ and $c_2 = 0.1$ as derived from time-of-flight measurement and other obtained parameters, the simulated velocity mapping image, a 2D projection of a whole ion cloud in laboratory frame, in Fig.7b turns out to agree satisfactorily with the experimental result in Fig. 7a. The best-fit image was achieved with $\alpha = 139°$ and $\chi = 166°$; the obtained angle $\chi$ is close to 180°, as arguable from the large value of the anisotropy parameter. On the other hand, the angle $\alpha$ deviates from 180°, which implies that the permanent dipole moment diverges from the direction of the recoil velocity vector. It is the asymmetric-top nature of 2-bromobutane that leads to the deviation of the angle $\alpha$ from 180°, which applies for a pure parallel transition of a symmetric top molecule. Further work should be devoted on the role of additional transitions, in the orientation probability distribution function in the photodissociation of this molecule [45-47]. A contribution from angle $\Omega$ is mapped to a radius on the image, and the outermost ring corresponds to $\Omega = 90°$. A near-ideal outermost ring ($\Omega = 90\pm3°$) gives rise to $\beta_1 = 0.35$ and $\beta_2 = 1.67$ in the simulation, which is in good agreement with the experimental results. It is noted that the experimental photofragment angular distributions (Fig. 7c) are averaged from the outermost 3 pixels, and thus the angles $\Omega = 90°$ with an uncertainty $\pm14°$ are taken into account in the area integration, as restricted within the image resolution. $\beta_1$ increases gradually and $\beta_2$ decreases, while counting the pixels toward the inner away from the outermost ring; for example, $\beta_1 = 0.39$, $\beta_2 = 1.40$ at $\Omega = 76°$, indicating the effect of the angle uncertainty on the experimental results.

## 4. Conclusions.

A 2-bromobutane supersonic beam was aligned and rotationally state-selected by an electrostatic hexapolar filter, whose rods are 70 cm long with a 4 mm radius. The rotational and vibrational temperatures were estimated to be 30 K and 300 K, respectively, by fitting the simulated focusing curve to the experimental one; the three conformers, T, G+ and G-, whose relative populations are 0.54, 0.28 and 0.18 respectively, gave a similar trend. The molecular orientation was evaluated by combining time-of-flight and slice-ion-imaging techniques. The results have shown that the orientation is achieved with a weak electric field, *ca.* 200 V/cm.

In order to study the photodissociation dynamics of the molecule, the traditional ion lens setup was modified. Specifically, a pulsed voltage that produces an orienting electric field of very short response time, 30 ns, was applied at the ion extraction stage. This improvement permits to change the intensity and the sense of the orienting field vector, which is constant in the traditional setup. Separation of the ion extraction and molecular orientation stages allows one to use strong electric field to orient molecules of a very high complexity, while maintaining a weak electric field for the ion extraction. It opens important perspectives in view of future applications.

Investigation by slice imaging on the photodissociation dynamics of 2-bromobutane at 234 nm, which yields Br ($^2P_{3/2}$) photofragment (we considered both the $^{79}$Br and the $^{81}$Br isotopes) at the ground state indicated as Br, and Br ($^2P_{1/2}$) at the first electronic excited state, indicate as Br*, permitted to shed light on the distribution of velocity and angular distribution of the photofragments reported in a previous publication. Zhou *et al.* [36] found a slow component of Br and Br* photofragments that was not detected by us. A detailed analysis by slice imaging technique has shown that it was given by the overlap of ion clouds of the isotopes $^{79}$Br and $^{81}$Br. Further, the photodissociation results of oriented 2-bromobutane yield the anisotropy parameters more complicated than the condition without any orientation and may have chance to look into the vector information associated with the photofragment.

**Supplementary Information.** The distribution of the selected rotational states for the three conformers of 2-bromobutane.

**Acknowledgements.**

F. P., A. L. and V. A. acknowledges the Italian Ministry for Education, University and Research, MIUR, for financial supporting: SIR 2014 "Scientific Independence for young Researchers" (RBSI14U3VF). F.P. acknowledges the Italian Ministry for Education, University and Research, MIUR, for financial supporting: FIRB 2013 "Futuro in Ricerca" (RBFR132WSM_003). Vincenzo Aquilanti thanks CAPES for the appointment as Professor Visitante Especial at Instituto de Fìsica, Universidade Federal de Bahia, Salvador (Brazil). V. A., A. L. and F. P. thank Professor K.-C. Lin for the several invitations to visit his group in Taipei, making possible the collaboration which eventually led to the preparation of this article. K.C.L. wishes to thank the financial support of Ministry of Science and Technology, Taiwan, under the contract no: NSC 102-2113-M-002-009-MY3.

**Bibliography.**

**Table 1.** Spectroscopic and electric properties of the three conformers of 2-bromobutane, T, G+ and G-: permanent dipole moments $d$ and their components $d_a$, $d_b$ and $d_c$ (in Debye); rotational constants A, B and C (in GHz); Ray's parameter; relative populations at vibrational temperature of 300 K.

| Conformer | $d$ (D) | $d_a$ (D) | $d_b$ (D) | $d_c$ (D) | A (GHz) | B (GHz) | C (GHz) | k | Relative population (300 K) |
|---|---|---|---|---|---|---|---|---|---|
| T | 2.80 | -2.71 | 0.63 | 0.32 | 6.38 | 1.54 | 1.30 | -0.91 | 0.58 |
| G+ | 2.70 | -2.22 | 1.47 | 0.35 | 3.88 | 2.20 | 1.49 | -0.41 | 0.28 |
| G- | 2.72 | -2.42 | 1.06 | 0.67 | 4.55 | 1.94 | 1.59 | -0.76 | 0.14 |

**Table 2.** The components of the moment of inertia $I_A$, $I_B$, $I_C$ (in a. m. u.), the kinematic invariants $\xi_1$, $\xi_2$, $\xi_3$, the hyperradius $\rho$ and the deformation indices $\xi_+$ and $\xi_-$ of the three conformers of 2-bromobutane.

| Conformer | $I_A$ | $I_B$ | $I_C$ | $\xi_1$ | $\xi_2$ | $\xi_3$ | $\rho$ | $\xi_+$ | $\xi_-$ |
|---|---|---|---|---|---|---|---|---|---|
| T | 283.05 | 1172.70 | 1389.18 | 2.88 | 1.35 | 0.49 | 3.22 | 0.15 | -0.63 |
| G+ | 465.44 | 820.89 | 1212.05 | 2.39 | 1.77 | 0.52 | 3.02 | 0.31 | -0.28 |
| G- | 396.90 | 930.89 | 1135.82 | 2.47 | 1.48 | 0.84 | 3.00 | 0.17 | -0.43 |

**Figures captions.**

**Figure 1.** The structure of 2-bromobutane and its three conformers, T, G+ and G-. In grey, we indicate the carbon atoms labeled from 1 to 4 according to the IUPAC nomenclature; in red, we report the bromine atom; the hydrogens are represented in white.

**Figure 2.** Description of the experimental apparatus. The excimer laser (Lambda Physik LPX-200) pumped by dye laser (Lambda Physik LPD-300) is frequency doubled by BBO crystal in the auto tracker (AT, Inrad Autotracker III) to get photolysis/ionization laser at ~234nm. The photolyzed sample, pure 2-bromobutane vapor is emitted from the pulsed valve (PV) to form molecular beam. The sample molecules in the beam interacts with hexapole (HP) electric field by Stark effect to be rotationally-selected and focused into the intersection region with the laser. The excimer laser and the pulsed valve are triggered by a digital delay/pulse generator (DG1, SRS DG-535) to synchronize them. The photofragment Br is immediately ionized to be mapped onto micro-channel plate and phosphor screen detector assembly (MCP&PS), then recorded by a charge-coupled device camera (CCD). At the same time, the intensity and time-of-flight are continuously monitored on an oscilloscope (OS) by using photo multiplier tube (PMT). The MCP is gated by short time high voltage pulses from the gate pulse generator (GP, Kentech Instruments Ltd.) to observe only the center slice of ion cloud spheres. Since the excimer laser has a slight firing jitter, typically 30ns, the gate pulse is triggered by the photo diode (PD) signal via the other delay/pulse generator (DG2, SRS DG-535) to avoid the jitter.

**Figure 3.** The focusing curve of the 2-bromobutane pure vapor. The black squares represent the experimental results in increasing with an interval of 0.5 kV, from 0 to 5 kV on the hexapole state selector. The continuous lines represent the simulated results. The red, blue and grey lines represent the focusing curves of the conformers T, G+ and G-, respectively. The black line indicates the simulation of the total focusing curve.

**Figure 4.** (a) Comparison of time-of-flight measurements with (red) and without (black) hexapole voltage. The REMPI signal of Br* at 234.0 nm with laser polarization is adjusted to be parallel to the time-of-flight axis. The background signals have already been subtracted. The upper image shows the ion cloud spheres of each isotope. The lighter $^{79}$Br is detected earlier, having partial overlap with the heavier $^{81}$Br. Panels (b1) and (b2) report the deconvolutions of the time-of-flight results when the hexapole voltage is not applied (b1) and applied (b2). The anisotropy parameter $\beta$, the velocity distribution of photofragments, the molecular beam velocity and its spread

were taken into account for the deconvolutions, although contribution of the beam velocity and of its spread is small. The red dashed line represents the contribution of $^{79}$Br and the blue dotted line represents that of $^{81}$Br. The green solid line is the sum of the two contributions.

**Figure 5.** (a) Time-of-flight signals with and without the pulsed orientation field. The hexapole voltage was kept at 4.0 kV for both cases. The black line is the result without the pulse and the red line is the result with the pulsed field which enhances the later peak for both isotopes. (b) Scheme of the application of the pulsed voltage in the extraction and orientation mode. R indicates the repeller and L1 indicates the first extractor lens plate.

**Figure 6.** The sliced images of Br and Br$^*$ photofragments from 2-bromobutane at 233.7 nm and 234.0 nm respectively. The velocity distributions and the angular distributions of each photofragment are in right panels. The polarization of photolysis laser was vertical in the figure plane. The gate pulse width was 25 ns to cut out a 10 % contribution of the $^{79}$Br ion cloud sphere.

**Figure 7.** (a) The ion image of Br* photofragment dissociated from 2-bromobutane. The laser polarization is tilted 45° toward the MCP surface. The image is acquired with a 400 ns long gate width in order to observe the whole ion sphere. (b) The simulated image of photofragments from oriented molecules with angles α and χ are 139° and 166° respectively. The colorbar at the side is shared in both (a) and (b). (c) The comparison of experimental and theoretical fitting of angular intensity of the outer ring. (d) Definition of the recoiling velocity vector **v**, the permanent dipole moment **d**, the transition dipole moment **μ**, the laser polarization **ε** and of the orienting field **O** in the molecular recoil frame *xyz*, where the *z*-axis corresponds to **v** and the *xz*-plane is defined by **v** and **ε**. Vectors are identified by two angles within parenthesis, the first one indicates the angle between the considered vector and the *z*-axis (or equivalently **v**), the second one is the angle between the vector and **ε**. (e) The geometry of experimental setup (HP indicates the hexapole) and the laser polarization in the laboratory frame *XYZ*, with the *Z*-axis parallel to the time-of-flight axis; the laser is radiated along the *Y* axis and its polarization is tilted by 45° in the *XZ*-plane. The vectors also in this case are identified by two angles (in the laboratory frame) in parenthesis: **ε** lies on the *XZ*-plane and *Γ* is the angle between **ε** and the *Z*-axis; **v** is defined by the elevation *Ω* and the azimuth *Θ*; finally, **O** depends on the elevation *Δ* and on the azimuth *Φ*.

**Figures.**

**Figure 1.**

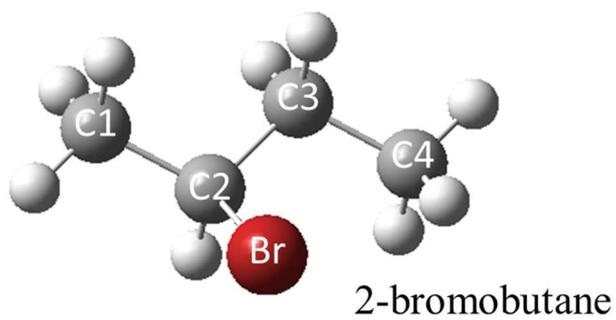

2-bromobutane

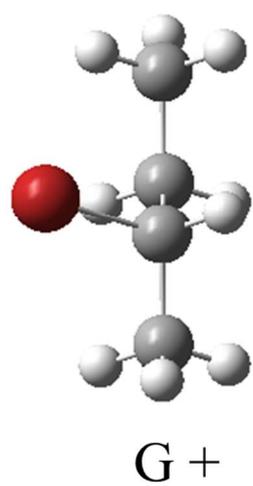

G +

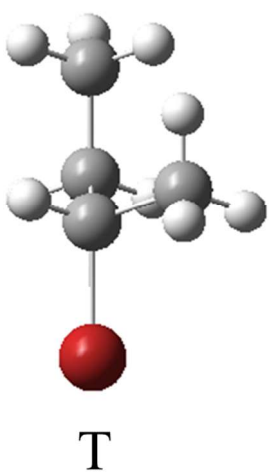

T

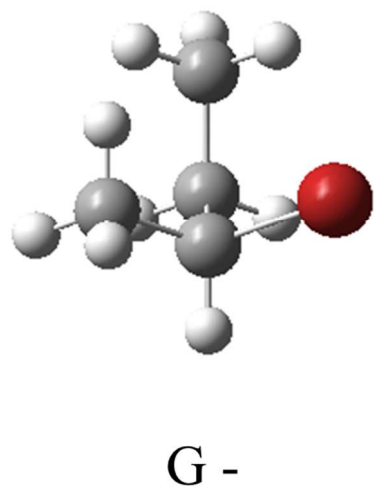

G -

**Figure 2.**

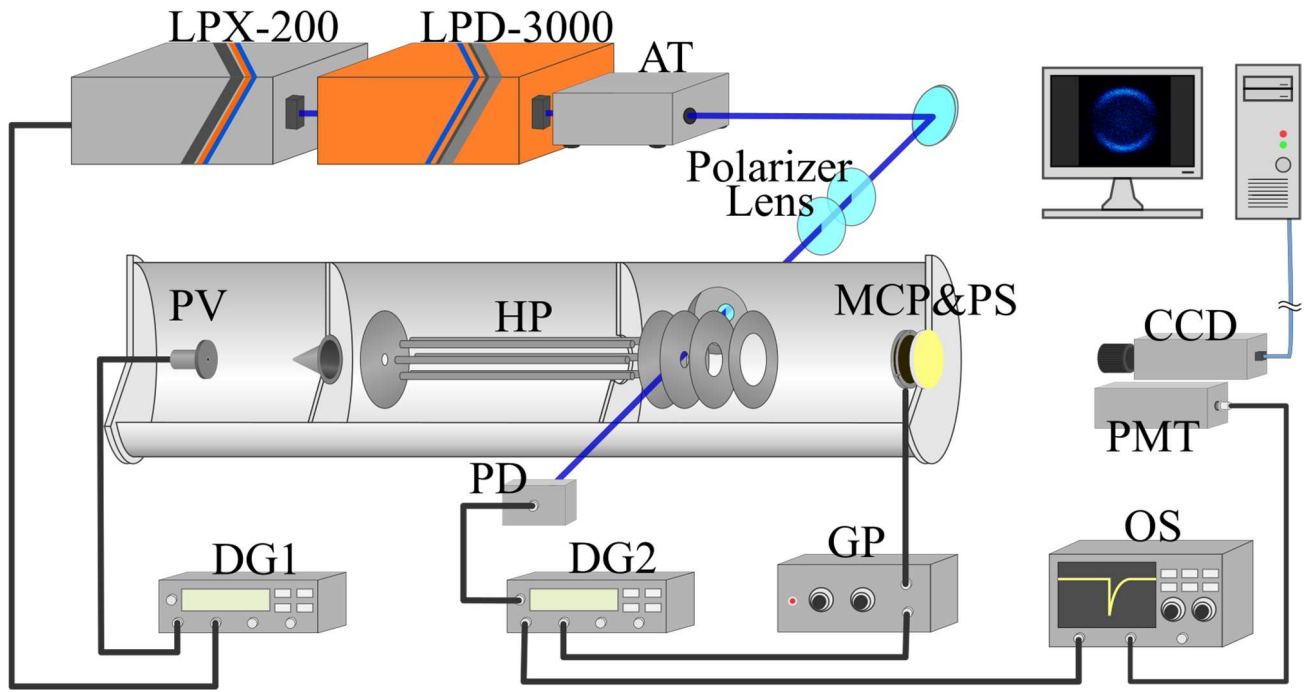

**Figure 3.**

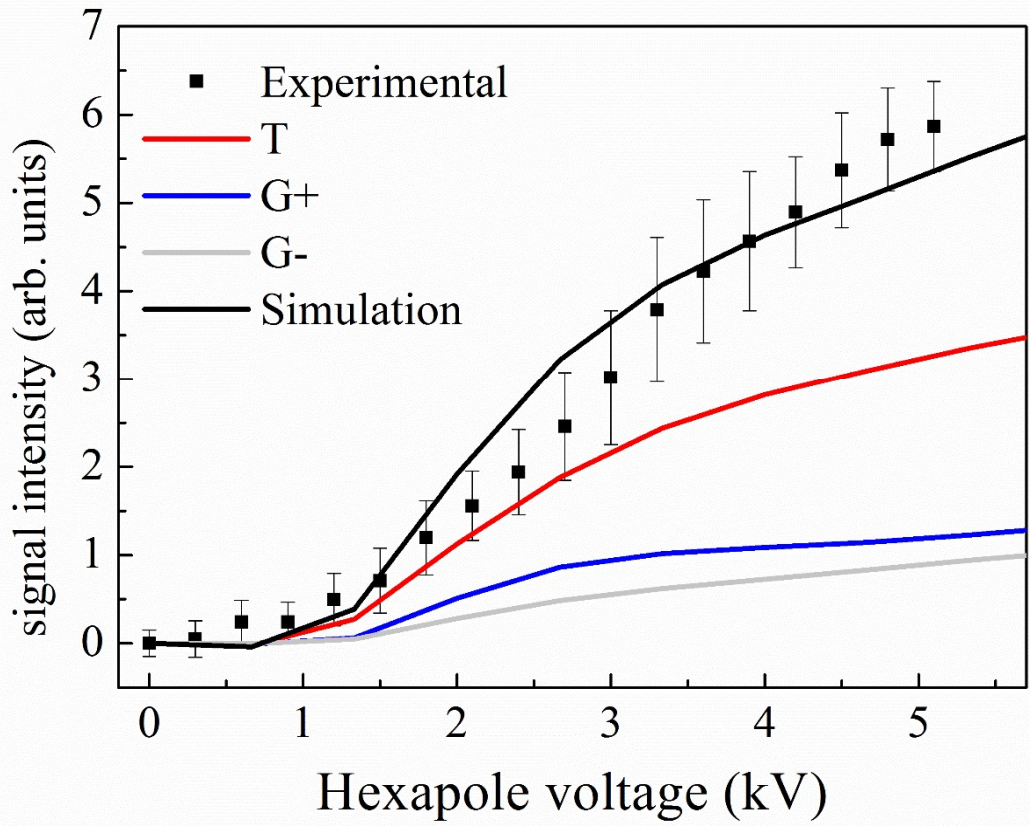

**Figure 4.**

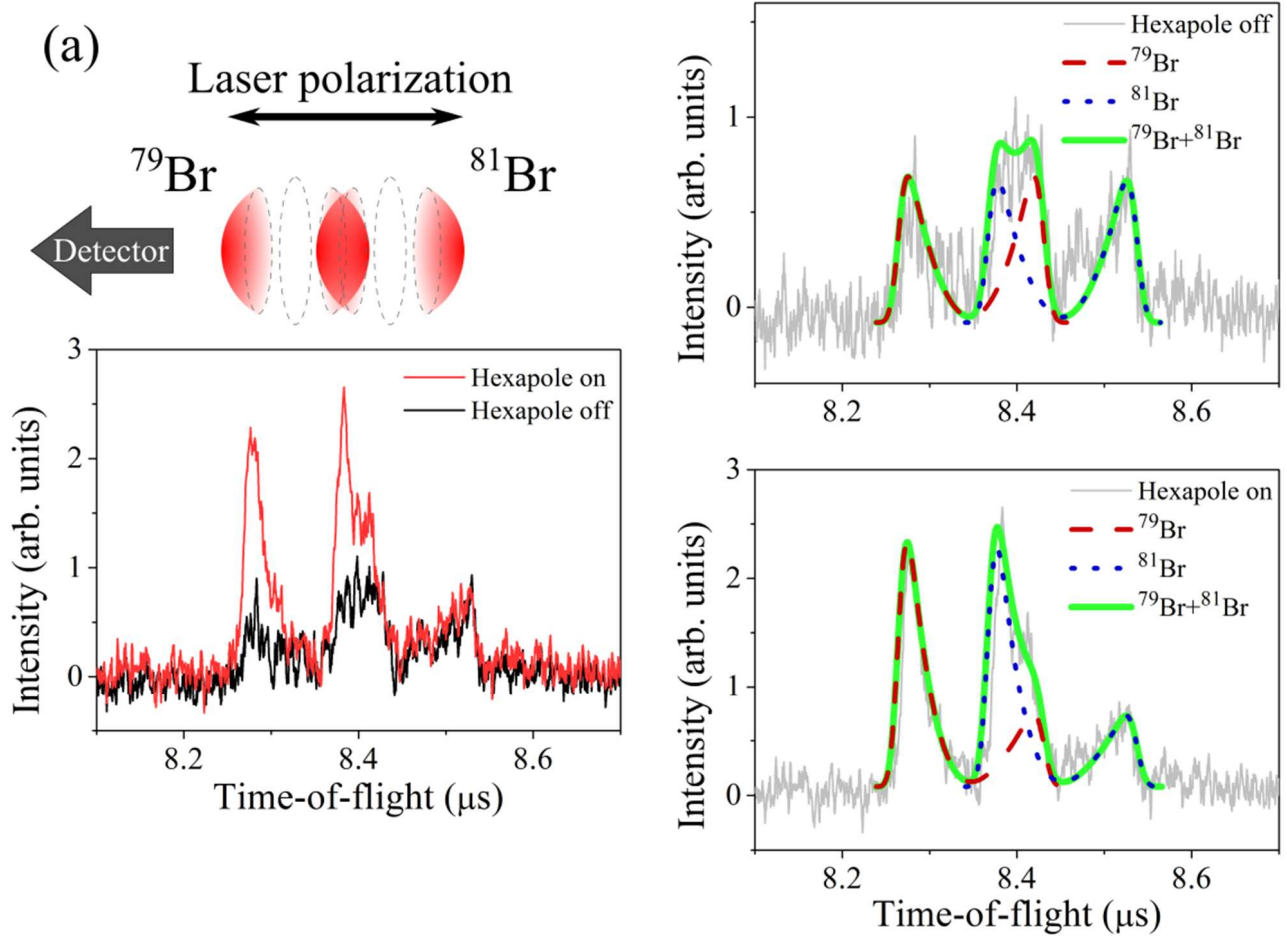

**Figure 5.**

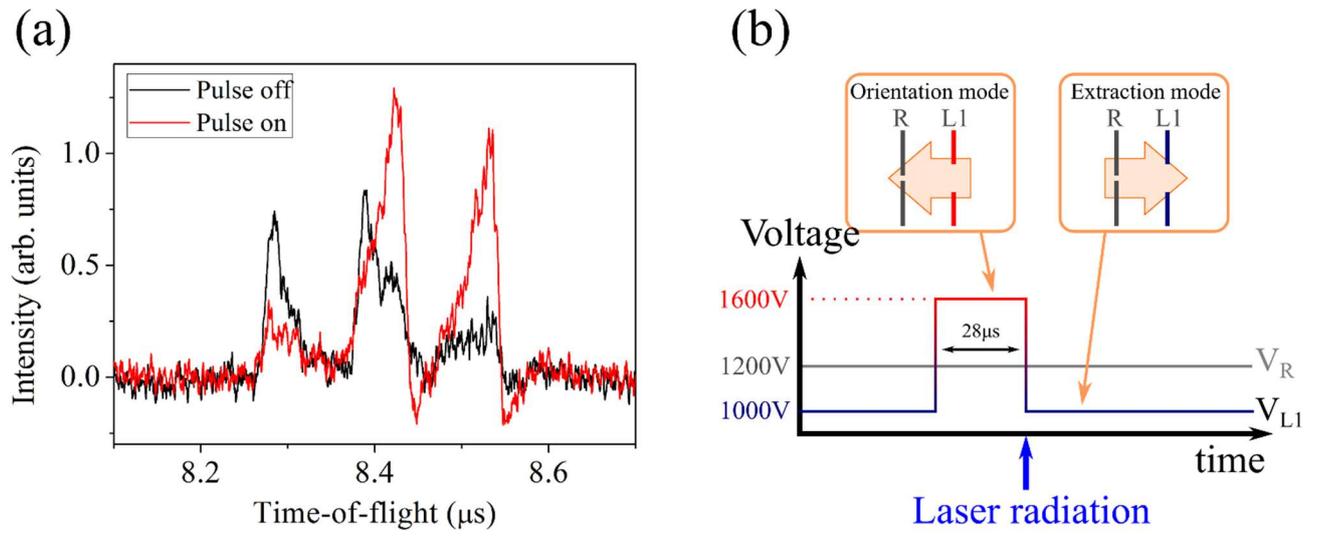

**Figure 6.**

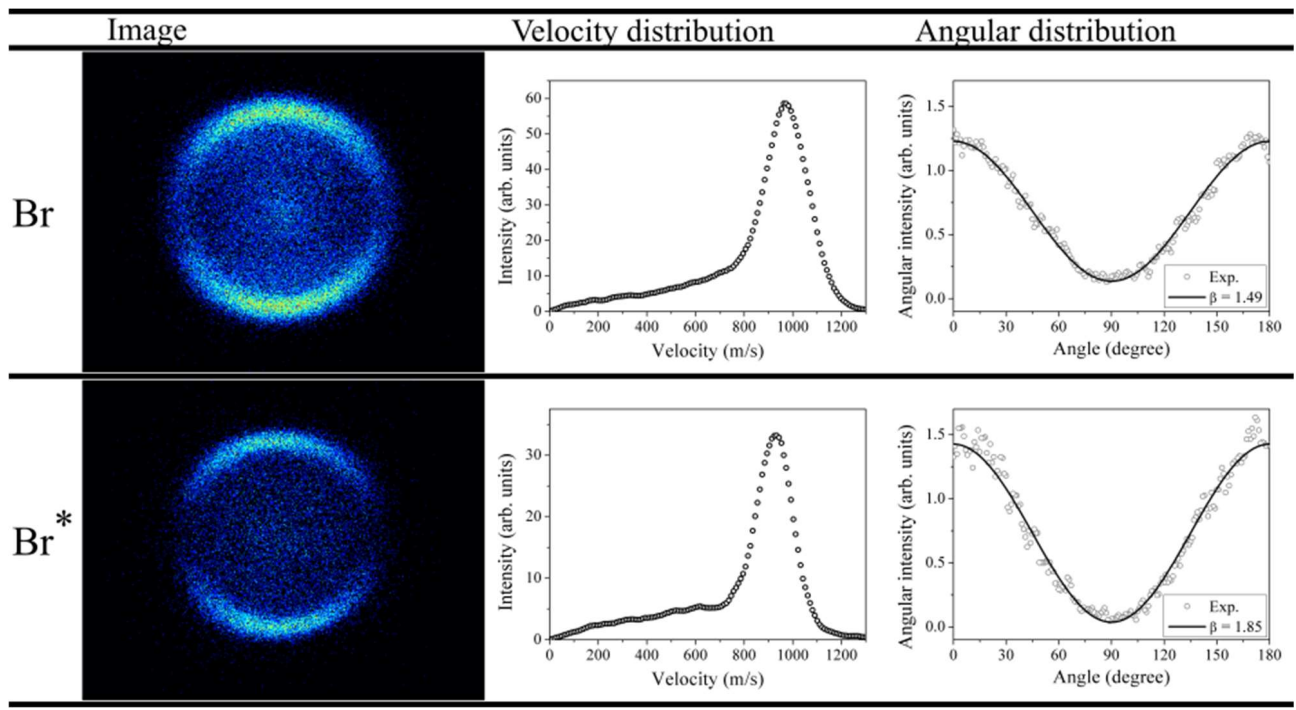

**Figure 7.**

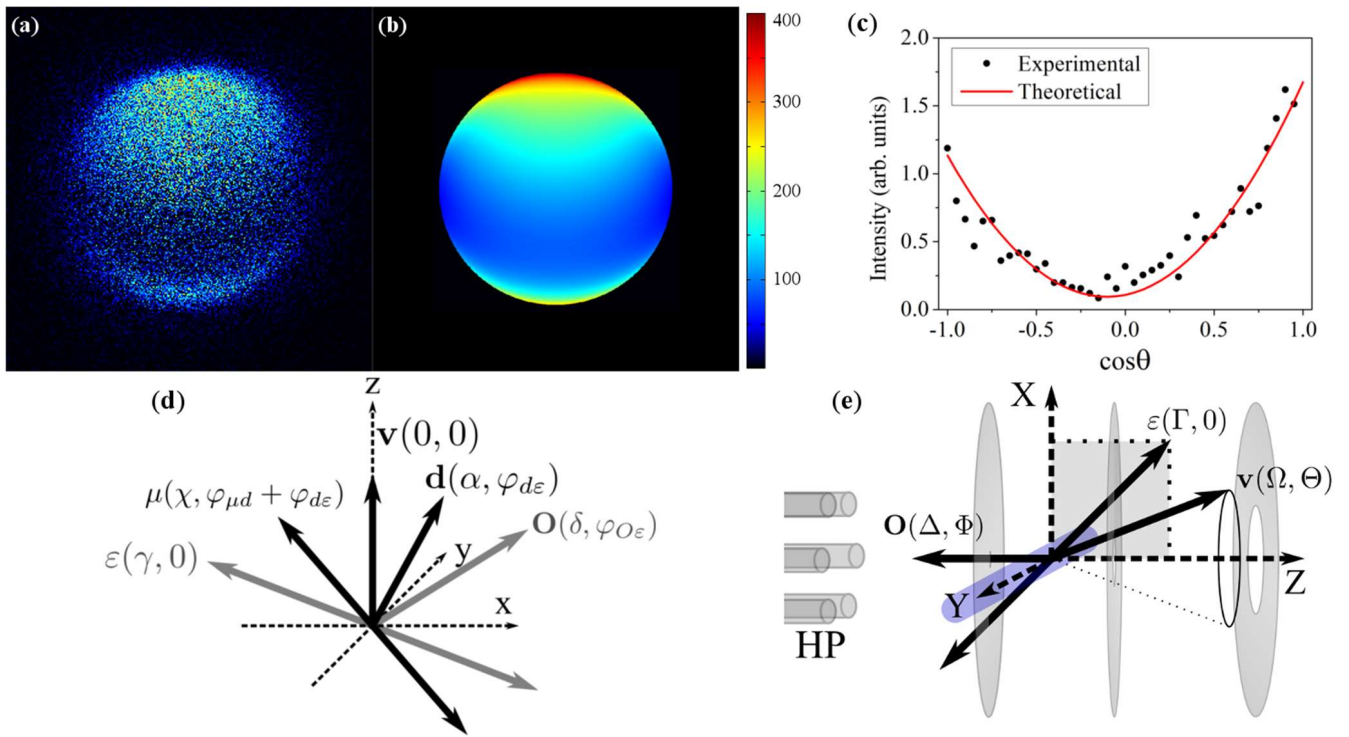